# Polarizability Versus Mobility: Atomistic Force Field for Ionic Liquids

Vitaly Chaban

University of Rochester, Rochester, New York 14627-0216, United States,
e-mail: v.chaban@rochester.edu

Based on classical molecular dynamics simulations, we discuss the impact of Coulombic interactions on a comprehensive set of properties of room temperature ionic liquids (RTILs) containing 1,3-dimethylimidazolium ($MMIM^+$), N-butylpyridinium ($BPY^+$), and bis(trifluoromethane sulfonyl)imide ($TFSI^-$) ions. Ionic transport is found to be noticeably hindered by the excessive Coulombic energy, originating from the neglect of electronic polarization in the condensed phase of these RTILs. Starting from the models, recently suggested by Lopes and Padua, we show that realistic ionic dynamics can be achieved by the uniform scaling of electrostatic charges on all interaction sites. The original model systematically overestimates density and heat of vaporization of RTILs. Since density linearly depends on charge scaling, it is possible to use it as a convenient beacon to promptly derive a correct scaling factor. Based on the simulations of [BPY][TFSI] and [MMIM][TFSI] over a wide temperature range, we conclude that suggested technique is feasible to greatly improve quality of the already existing non-polarizable FFs for RTILs.

**Introduction**

Room-temperature ionic liquids (RTILs) are a broad class of ionic compounds, which are formed by large asymmetric organic cation and organic or inorganic anion[1-9]. Lots of organic cations can be utilized including imidazolium$^+$, pyridinium$^+$, pyrrolidinium$^+$, tetraalkylammonium$^+$, tetraalkylphosphonium$^+$, triazolium trialkylsulfonium$^+$, piperidinium$^+$, pyrazolinium$^+$, guanidinium$^+$, etc. In turn, the most common anions are tetrafluoroborate$^-$, alkylsulfonate$^-$, dicyanamide$^-$, chloride$^-$, bromide$^-$, hexafluorophosphate$^-$, bis(trifluoromethylsulfonyl)imide$^-$, and acetate$^-$. The total number of RTILs, synthesized so far, exceeds 500 and still increases. From the practical point of view, RTILs can boast a wide variety of potential applications[7,8], not limited to components of robust electrolytes for electrochemical devices[10,11], reaction medias[12], lubricants[13], solvents for synthetic and catalytic schemes[14,15].

The prominent properties of RTILs, as salts with very low melting points, are determined by the structure and inter-ionic interactions. In addition to the interactions known for conventional organic solvents, such as hydrogen bonding, dipole–dipole and van der Waals interactions, ionic liquids exhibit strong ionic interactions, which make them extremely miscible with all polar compounds[16]. Meanwhile, alkyl chains on the cations favor their solubility in weakly polar liquids. Hydrogen bonding is believed to exist between oxygen and/or halide atoms of the anion and hydrogen atoms on the imidazolium or pyridinium ring of the cation[17].

In line with today's fashion in chemistry, probably the most exciting feature of RTILs is an ability to design a liquid with a set of specific properties. A particular liquid for each specific application is expected to be engineered. Therefore, RTILs are sometimes called "designer solvents"[18]. Precise tuning of physical and chemical properties is straightforward by variation of the length and branching of the cation's alkyl groups, as well as shape and polarity of the anion. For instance, the change of length of the 1-alkyl chain from 1 to 9 on 1-alkyl-3-methyl-imidazolium hexafluorophosphate can turn the liquid from being water-soluble to water-insoluble[16]. With all these considerations in mind, computer simulations gain extreme importance for RTILs, since they are the cheapest way to derive nearly all relevant physical properties for any coveted composition.

Although atomistic simulations of RTILs[19-31] were intensively carried out during the last decade, there is still a significant problem associated with most of the existing force fields (FF)[32-47]. All non-polarizable FFs for classical molecular dynamics (MD) simulation, derived using the conventional procedures with electrostatic charges on the interaction sites of cations and anions computed separately, greatly underestimate ionic dynamics. This results in up to 10 times lower diffusion constants and ionic conductivity as compared to the experimental values. As a result, an unreasonably high shear viscosity is observed in such simulations (~1000 cP). At the same time, structure and thermodynamics properties of RTILs are reproduced much better (generally, the deviations are within 3%).

Certain efforts have been recently made to improve the situation, including the decrease of electrostatic charges[48,49], fine tuning both sigmas and epsilons for selected interaction sites[50,51], tuning bonded interactions to reproduce newly obtained experimental properties[46]. Unfortunately, most of these refinements are quite artificial and do not have a strong physical background. The common goal of all these attempts is to weaken interactions, particularly, Coulombic forces. Consequently, they work only for a particular RTIL and are not portable to another species. The best solution so far is proposed by Borodin[52] within the framework of many-body polarizable force field. The electronic polarization effects, taking place in condensed phases of ionic liquids, are incorporated *via* an isotropic atomic dipole formalism using Thole screening. In turn, the electrostatic potential around molecules is described *via* permanent charges located on atoms and off-atom sites. The undoubtful success of this FF[19,20,52-55] proves that ionic transport in RTILs is driven by electronic polarization effects. These effects should be thoroughly taken into account, while a new FF is being developed.

The only con of the Borodin's FF is its noticeably higher computational cost as compared to the conventional non-polarizable ones. It can be sensible, since RTILs often require extensive simulations.

Recently, we reported a procedure whose purpose is to tune the simulated electrostatic potential[56]. It is based on uniform scaling of the partial electrostatic charges on all interaction sites. The uniform scaling needs a scaling factor, and it was derived from the electrostatic potential computed using density functional theory. The explicit ionic environment was utilized in order to reproduce realistic local structure of RTILs. Importantly, the uniform scaling of electrostatic charges was shown to be an adequate approximation in case of bulk imidazolium-based ionic liquids. This finding opens new avenues to obtain reasonable Coulombic interaction energy and observe realistic dynamics within the framework of classical MD studies.

In the present work, we investigation of general relationships between Coulombic energy and various physical properties of the simulated system is carried out. Heat of vaporization, specific density, radial distribution function (RDF), cumulative numbers (CNs), diffusion constant ($D$), shear viscosity ($\eta$) and ionic conductivity ($\sigma$) of N-butylpyridinium bis(trifluoromethane sulfonyl)imide ([BPY][TFSI]) and 1,3-dimethylimidazolium bis(trifluoromethane sulfonyl)imide ([MMIM][TFSI]), representing pyridinium-based and imidazolium-based RTILs, are reported. The optimized geometries of $BPY^+$, $MMIM^+$, and $TFSI^-$ are depicted in Figure 1. Based on the obtained results, all properties, simulated over a wide temperature range, are close to their experimental values at similar scaling factors. The best-performing scaling factors range between 0.7 and 0.8, corresponding to the Coulombic energy decrease by 36-51%. In addition, we demonstrate that density of RTIL can be used as a convenient beacon to determine this scaling factor. The suggested technique minimizes computational efforts needed to refine the already existing FFs.

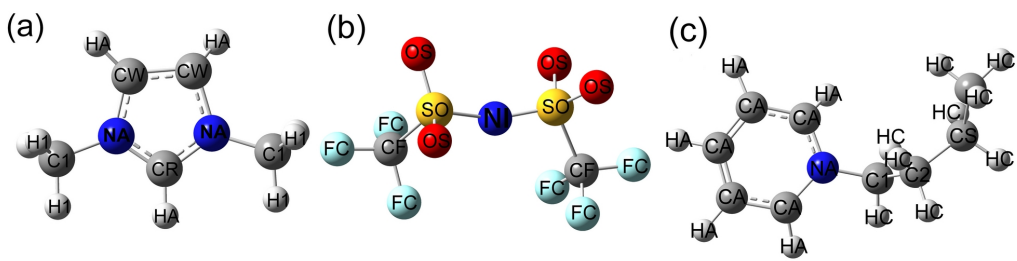

Figure 1. The optimized geometries of the simulated ions, (a) N-butylpyridimium$^+$ ($BPY^+$), (b) 1,3-dimethylimidazolium$^+$ ($MMIM^+$), (c) bis(trifluoromethane sulfonyl)imide$^-$ ($TFSI^-$). The optimization was carried out using density functional theory computations with B3LYP functional and 6-311++G** basis set.

**Computational Setup**

In order to compute density, heat of vaporization, radial distribution function, cumulative numbers, diffusion constant, shear viscosity and ionic conductivity of RTILs, we derive phase trajectories using classical molecular dynamics with fixed pairwise interaction potentials. All atoms of N-butylpyridimium bis(trifluoromethane sulfonyl)imide and 1,3-dimethylimidazolium bis(trifluoromethane sulfonyl)imide, including hydrogens, are represented as separate interaction centers possessing Lennard-Jones (12,6) parameters and electrostatic charges. All MD runs are carried out using the GROMACS 4.0 simulation package[57] in the constant temperature and constant pressure ensemble. Constant temperature of 303, 353 and 403 K (for separate simulations) is maintained using the novel V-rescale thermostat[58] with a response time of 1.0 ps. The constant pressure (1 bar) is maintained using Parrinello-Rahman barostat[59] with a relaxation time of 4.0 ps. The conventional leap-frog algorithm is used to integrate the equations of motion with a time-step of 0.001 ps at all simulated temperatures. The hindered ionic dynamics of the investigated RTILs requires long equilibration and production stages of MD simulations. In the present study, we use 10,000 ps for equilibration and from three to five consequent runs of 25,000 ps for production stage. To ensure extensive trajectory sampling, each productive run begins with an assignment of random velocities to ions, corresponding to the simulated temperature (303, 353 or 403 K).

The simulated systems of two RTILs, consisting of 128 ion pairs each, are placed in cubic MD boxes with periodic boundary conditions, applied along all Cartesian directions. The long-range electrostatic forces are treated by the well-known Ewald summation method with the cut-off distance for real-space component equal to 1.3 nm. The Lennard-Jones part is treated by the shifted force method with a switch region between 1.1 nm and 1.2 nm. The list of neighbors is updated every 0.02 ps (10 time-steps) within a sphere of radius of 1.3 nm.

The research is started with FF models proposed by Lopes and Padua[36,47]. The electrostatic potentials are gradually changed by uniform decrease of Coulombic charges on all interaction sites of counterions. Overall, six sets of charges, $q_i$, are considered, $1.0q_i$, $0.9q_i$, $0.8q_i$, $0.7q_i$, $0.6q_i$ and $0.5q_i$, where $q_i$ is an original value published by Lopes and Padua[36,47]. Since Coulombic energy is proportional to the square of charge $q_i$, its decrease corresponding to such scaling procedure is 0, 19, 36, 51, 64 and 75%, respectively. In turn, neither Lennard-Jones, nor intra-molecular parameters of interaction are modified in this work.

We derive distribution functions, heats of vaporization, transport properties (diffusion constant, shear viscosity and ionic conductivity), writing down the atomic coordinates and interaction energies every 0.02 ps (10 time-steps) at 403 K, every 0.03 ps (15 time-steps) at 353 K and every 0.04 ps (every 20 time-steps) at 303 K. Diffusion constants are derived on the consequent parts of trajectory of 10,000 ps, whereas shear viscosity and ionic conductivity require longer times (25,000 ps) to obtain reliable values. The convergence of all transport properties is assured by calculating them as a function of simulation time. A few consequent runs with randomly generated velocities, as mentioned above, are applied to estimate the averages and standard deviations of the resulting values. In turn, structure pair colleration functions, specific density and heat of vaporization can be derived using significantly poorer sampling (1,000 ps), provided that the simulated systems are equilibrated thoroughly.

Density and its fluctuation of the RTIL systems (Figure 2) during the simulation are estimated from the oscillations of the MD box volume. Generally speaking, these oscillations depend on both barostat parameters and size of the simulated system. So, it is generally possible to decrease them, if needed. Diffusion constant (Figure 3), $D$, is computed *via* the Einstein relation, through plotting mean square displacements of all atoms of each species. Shear viscosity (Figure 4), $\eta$, is obtained by integrating the autocorrelation function of the off-diagonal elements of pressure tensor. This method allows to get viscosity of the system using equilibrium MD methodology, although pretty long trajectories are required. Ionic conductivity (Figure 6), $\sigma$, is obtained in the framework of the Einstein-Helfand formalism from the linear slope of mean-square displacements of the collective traslational dipole moment. Importantly, for this method to work correctly, periodic boundary conditions should be removed prior to computation, i.e. "free" diffusion is necessary. Radial distribution function (Figure 7), $g_{ij}(r)$, is calculated using its classical definition on the trajectory parts of 1,000 ps. Corresponding cumulative numbers (Figure 7), $n_{ij}(r)$, are computed as integrals of $g_{ij}(r)$ taken from 0 to $r$. Heat of vaporization (Figure 8), $H_{vap}$, is estimated at 303 K only. Importantly, vapor phase of RTILs is assumed to contain only neutral ionic pairs, [MMIM][TFSI] and [BPY][TFSI], which do not interact with one another.

**Results and Discussion**

The specific density of [MMIM][TFSI] and [BPY][TFSI] at 303, 353 and 403 K as a function of the electrostatic scaling factor is depicted in Figure 2. Calculated using the original FF[36,42], density is systematically overestimated. With a scaling factor decrease, density also decreases linearly with a slope equal to 192, 221, 249 kg/m$^3$ for 303, 353, 403 K, respectively. Remarkably, the slope increases as temperature increases. Importantly, at all temperatures the simulated density approaches its experimental value at approximately the same scaling factor (~0.70 for [MMIM][TFSI] and ~0.76 at [BPY][TFSI]). Therefore, one can refine these models for density using the uniform decrease of Coulombic interaction energy (i.e. charges) only. This finding about the success of uniform scaling well agrees with our previous studies of 1-ethyl-3-methylimidazolium tetrafluoroborate and 1-butyl-3-methylimidazolium tetrafluoroborate simulated over a wide temperature interval[56]. Although it may contradict intuition, the impact of electrostatics on the densities of RTILs appears to be uncritical. For [MMIM][TFSI], the withdrawal of 75% of Coulombic energy results only in 6-8% decrease of specific density. Thus, both RTILs still remain quite dense liquids, much heavier than water. As temperature grows from 303 to 403 K, the relative decrease of density due to scaling enlarges from 6 to 8%.

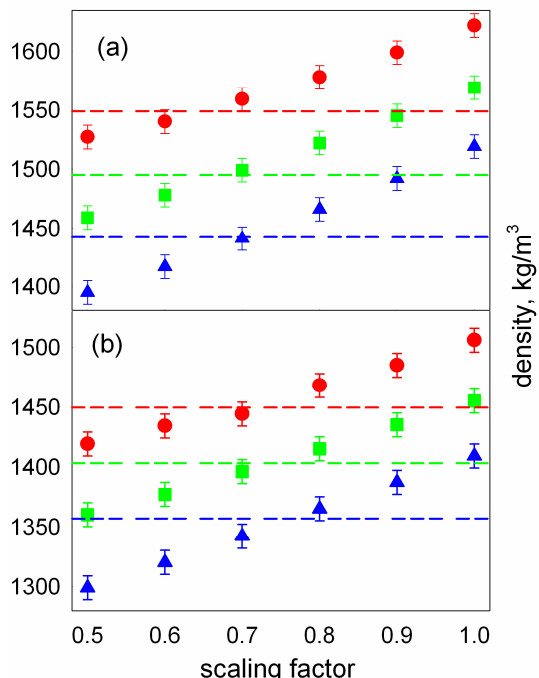

Figure 2. The simulated densities of (a) [MMIM][TFSI] and (b) [BPY][TFSI] with the scaling factors from 1.0 down to 0.5 at 303 (red circles), 353 (green squares) and 403 K (blue triangles). The dashed horizontal lines of the same color indicate experimental values at the corresponding temperatures. The standard deviation of the simulated density is 0.6-0.8%.

As well as the majority of other FFs, Lopes' FF greatly underestimates cationic and anionic diffusion constants of [MMIM][TFSI] and [BPY][TFSI] (Figure 3). The values derived from the 10 ns MD simulations of [MMIM][TFSI] at 303, 353, and 403 K are (in $10^{-11}$ m$^2$/s) 0.42, 2.8, 13 for [MMIM]$^+$, and 0.38, 2.1, 7 for [TFSI]$^-$, whereas experimental $D$ are 5.2, 22, 48 for cation, and 3.2, 13, 30 for anion. For [BPY][TFSI], the situation is pretty similar, i.e. the simulated $D$s are 0.6, 3.0, 10 (in $10^{-11}$ m$^2$/s) for [BPY]$^+$ and 0.4, 2.3, 8.3 for [TFSI]$^-$, whereas experimental $D$s are 2.7, 14, 36 for cation, and 2.2, 11, 30 for anion. Noticeably, at 303 K the simulated and experimental $D$s differ by one order of magnitude for [MMIM][TFSI] and by a half of an order for [BPY][TFSI]. However, as temperature increases to 403 K, this difference decreases to about four times for both RTILs. Unlike density, the relationship between diffusion and scaling factor (or Coulombic energy decrease) is much more pronounced and can be well depicted by $D = a \cdot \exp(-b \cdot f_{sc})$ (Figure 5a), where $f_{sc}$ is a scaling factor, and $a$ and $b$ are arbitrary constants. Applying $f_{sc}$ to initial charges allows to approach experimental diffusion constants for both RTILs. The successful scaling factors are 0.72-0.74 for [MMIM][TFSI], and 0.76-0.78 for [BPY][TFSI]. Importantly, $f_{sc}$ is nearly the same (Table 1) for all three temperatures (303 K, 353 K and 403 K). If $f_{sc}$ decreases to 0.5, it leads to a dramatic increase of $D$ for all simulated species. However, even losing 75% of their electrostatic interaction energy, RTILs are liquids with a very slow particle dynamics. For instance, at room temperature $D$ is less than $0.3 \times 10^{-11}$ m$^2$/s, which is much smaller than for conventional molecular liquids (~1-3×10$^{-11}$ m$^2$/s).

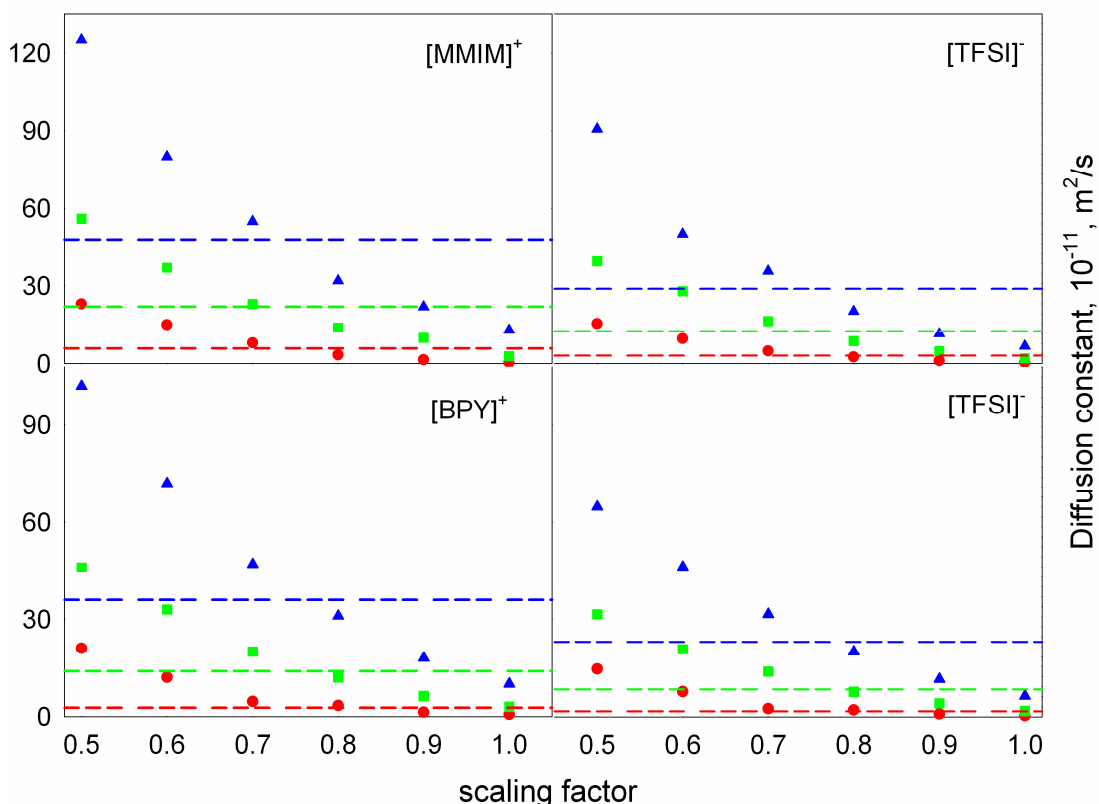

Figure 3. The simulated diffusion constants of [MMIM][TFSI] and [BPY][TFSI] with the scaling factors from 1.0 down to 0.5 at 303 (red circles), 353 (green squares) and 403 K (blue triangles). The dashed horizontal lines of the same color indicate experimental values at the corresponding temperatures. The standard deviation of the simulated diffusion constant is 20-30%.

According to the Stokes-Einstein relationship, self-diffusivity and shear viscosity of the same system are inversely proportional. Expectedly, the hindered ionic dynamics, characteristic to original FF, leads to appropriately overestimated viscosity (Figure 4). The most critical deviations from experiment are observed at room temperature. Again, the uniform scaling of charges on both cation and anion deals with this problem successfully. The best-performing $f_{sc}$s are 0.74-0.76 and 0.78 for [MMIM][TFSI] and [BPY][TFSI], respectively. The dependence of viscosity upon the scaling factor is exponential at all temperatures (Figure 5b) with a correlation coefficient of more than 98%. The successful scaling factors are almost the same for density, diffusion and viscosity of these RTILs. It should be noted that the Einstein method for viscosity converges very slowly and usually provides a large standard deviation of the resulting value. Therefore, a few consequent simulations originating from different starting points of phase space are required. Larger MD systems can improve this situation, although at a considerably higher computational cost.

Table 1. The best-performing scaling factors for specific density, heat of vaporization, diffusion constants, shear viscosity and ionic conductivity of [MMIM][TFSI] and [BPY][TFSI], obtained through interpolation of the simulated data

|  | 303 K | 353 K | 403 K |
|---|---|---|---|
| Specific density | | | |
| [MMIM][TFSI] | 0.68 | 0.70 | 0.70 |
| [BPY][TFSI] | 0.76 | 0.76 | 0.78 |
| Heat of vaporization | | | |
| [MMIM][TFSI] | 0.74 | — | — |
| [BPY][TFSI] | 0.70 | — | — |
| Cation diffusion constant | | | |
| [MMIM][TFSI] | 0.73 | 0.72 | 0.72 |
| [BPY][TFSI] | 0.78 | 0.76 | 0.76 |
| Anion diffusion constant | | | |
| [MMIM][TFSI] | 0.74 | 0.74 | 0.74 |
| [BPY][TFSI] | 0.78 | 0.78 | 0.76 |
| Shear viscosity | | | |
| [MMIM][TFSI] | 0.76 | 0.74 | 0.74 |
| [BPY][TFSI] | 0.78 | 0.78 | 0.78 |
| Ionic conductivity | | | |
| [MMIM][TFSI] | 0.64 | — | — |
| [BPY][TFSI] | 0.78 | 0.76 | 0.74 |

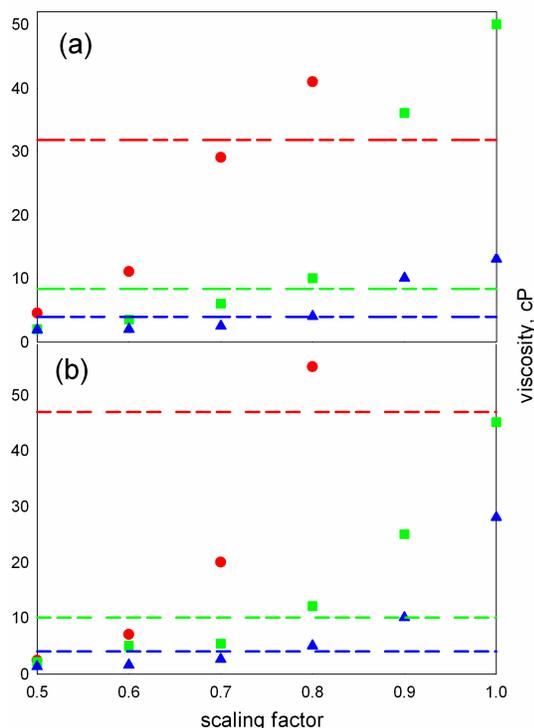

Figure 4. The simulated shear viscosities of (a) [MMIM][TFSI] and (b) [BPY][TFSI] with the scaling factors from 1.0 down to 0.5 at 303 (red circles), 353 (green squares) and 403 K (blue triangles). The dashed horizontal lines of the same color indicate experimental values at the corresponding temperatures. The simulated viscosities for [MMIM][TFSI] and [BPY][TFSI] with scaling factors of 1.0 and 0.9 at 303 K do not fit on the plot. Their values (in cP) are 370, 99, 300 and 140, respectively. The standard deviation of the simulated shear viscosity is around 40%.

Similarly to other properties under consideration, ionic conductivity, $\sigma$, (Figure 6) needs interaction energy to be decreased in order to reproduce experimental values. For [BPY][TFSI], this can be achieved with $f_{sc}$ equal to 0.74-0.78. The successful $f_{sc}$ is again in good agreement with obtained for density, diffusion constants and shear viscosity (0.76-0.78). However, in the case of [MMIM][TFSI] at 353 and 403 K, the values of $\sigma$ reported in Ref.[60] can not be approached, in spite of the scaling factor used. The simulated values are permanently smaller than in experiment, and deviation is equal to 30%. Probably, the only possible explanation of this inconsistency is connected with unreliable experimental values for [MMIM][TFSI]. This statement is based on the following theoretical considerations. [MMIM][TFSI] and [BPY][TFSI] are quite similar in the shape of ions, molecular mass and chemical structure. At 403 K, the experimental viscosities of these RTILs[60] are equal (4 cP), and the average diffusion constants are 39 and 34×$10^{-11}$ m$^2$/s[60], i.e. differ by only 13%. With these facts in mind, the difference of 43% in ionic conductivity looks really weird. Moreover, it was shown[56] that uniform scaling of charges for imidazolium-based RTILs provides realistic $\sigma$. Based on the current simulation, ionic conductivity of [MMIM][TFSI] should be expected around 4.0 and 2.0 S/m at 403 and 353 K, respectively. In the latter case, this result is consistent with all other properties of [MMIM][TFSI]. In particular, the difference in $\sigma$ between [MMIM][TFSI] and [BPY][TFSI] becomes 15-20% (at 353-403 K) instead of 43%. Unlike density, diffusion and viscosity, $\sigma$ reaches its maximum at $f_{sc}$=0.7 and does not vary as $f_{sc}$ goes down to 0.5. In other words, the acceleration of ionic motion due to the potential energy decrease is equally compensated by the decrease of partial Coulombic charges on each carrier of the current.

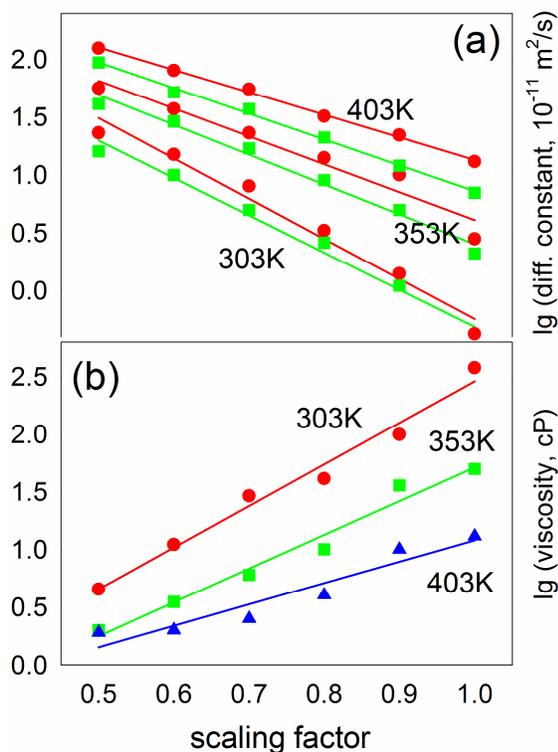

Figure 5. The logarithms of the simulated (a) diffusion constants and (b) shear viscosities of [MMIM][TFSI] at 303, 353, and 403 K. The cationic diffusion constants are depicted using red circles, whereas green squares show the anionic diffusions constants.

Although the FF of Lopes[36,42] fails to reproduce transport constants for RTILs, it provides reliable structure. Figure 7 depicts the radial distribution functions, $g_{HO}(r)$, and corresponding cumulative numbers, $n_{HO}(r)$, calculated between the hydrogen atom of the imidazolium ring and the oxygen atom of the anion for a series of $f_{sc}$ ranging from 1.0 down to 0.5. Importantly, even the electrostatic interaction energy decrease of 75% does not break structures of these RTILs. So, RDF approves the conclusion based on the analysis of specific density (Figure 2). Both maxima and minima of $g_{HO}(r)$ remain at their original positions, although the height of first maximum evidently decreases from ~2 to ~1.5 for both RTILs, as the scaling factor goes down to 0.5. Generally, it is possible to speak about very weak hydrogen bonding between H and O, which is ~20-25% longer than the conventional one (197 pm). Without pretending to judge about the impact of this hydrogen bonding on the structure of these RTILs, it should be noted that the corresponding interaction becomes somewhat weaker, if the scaling of any magnitude is applied. A more detailed investigation of the oxygen-hydrogen interaction may be important in order to understand its influence on the ionic mobility in these ionic liquids.

No significant change in local structure is also indicated by $n_{HO}(r)$ (Figure 7). It shows the co-ordination numbers of [MMIM]$^+$ and [BPY]$^+$ with respect to [TFSI]$^-$ based on the current distance between hydrogen atoms of imidazolium ring of the cation and oxygen atoms of the anion. Since [TFSI]$^-$ contains four oxygen atoms, the value of $n_{HO}(r)$ at given $r$ should be divided by this number to obtain CN for cation as a whole. The first co-ordination sphere (corresponding to the first maximum on the RDFs) of each cation consists of just one anion, both in [MMIM][TFSI] and [BPY][TFSI] RTILs. In turn, the second shell contains four anions ($n_{HO}\approx16$), and this CN does not noticeably vary with a scaling factor. To recapitulate, the analysis of RDFs and RCNs shows that the decrease of $f_{sc}$ from 1.0 to 0.5 does not break the positions of peaks and distribution of the co-ordination numbers, although the height of the peaks becomes smaller. This underlines that the structure of RTILs softens, leading to a larger ionic mobility (Figures 3-5).

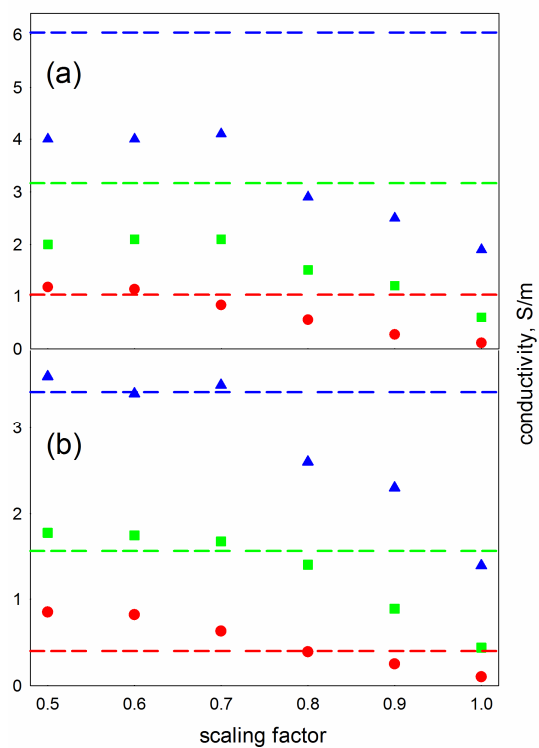

Figure 6. The simulated ionic conductivities of (a) [MMIM][TFSI] and (b) [BPY][TFSI] with the scaling factors from 1.0 down to 0.5 at 303 (red circles), 353 (green squares) and 403 K (blue triangles). The dashed horizontal lines of the same color indicate experimental values at the corresponding temperatures. The standard deviation of the simulated ionic conductivity is around 40%.

The heat of vaporization, $H^{vap}$, is probably the most common thermodynamics property to calibrate the phenomenological force fields. It directly reflects the physical forces and energies which keep the condensed matter system as a unit. So far, the experimental $H^{vap}$, although often estimated from indirect measurements, are available for most popular RTILs. The unmodified Lopes' FFs[34,36,38,41,42,44,45,47] systematically overestimate $H^{vap}$ (Figure 8). In turn, the decrease of the Coulombic interaction energy, achieved by the scaling procedure, naturally decreases $H^{vap}$. The experimental values of $H^{vap}$ are approached at $f_{sc}$=0.74 and $f_{sc}$=0.70 for [MMIM][TFSI] and [BPY][TFSI], respectively. Remarkably, the relationship between [MMIM][TFSI] and [BPY][TFSI] is reproduced at any scaling factor. To summarize, the best-performing $f_{sc}$ is nearly the same for thermodynamics and transport properties, whereas structure correlation functions are generally insensitive to it.

It may be interesting to estimate the overall impact of the Coulombic interactions on the ionic motion of the RTILs. Such kind of study was previously reported for N-methyl-N-propylpyrrolidinium (MPPY) bis(trifluoromethanesilfonyl)imide[61] over the temperature interval between 303 and 393 K. All electrostatics interactions were simultaneously turned off resulting in the neutral and non-polar MPPY and TFSI molecules. It was found that diffusion constants increased by 19-42 times, while the shear viscosity was 16-42 times lower. A significantly more pronounced effect was observed at low temperature. It is consistent with a general fact that electronic polarization dominates only at relatively low temperatures. Also expectedly, no long time correlations were found for the motion of this type of artificially neutralized molecules.

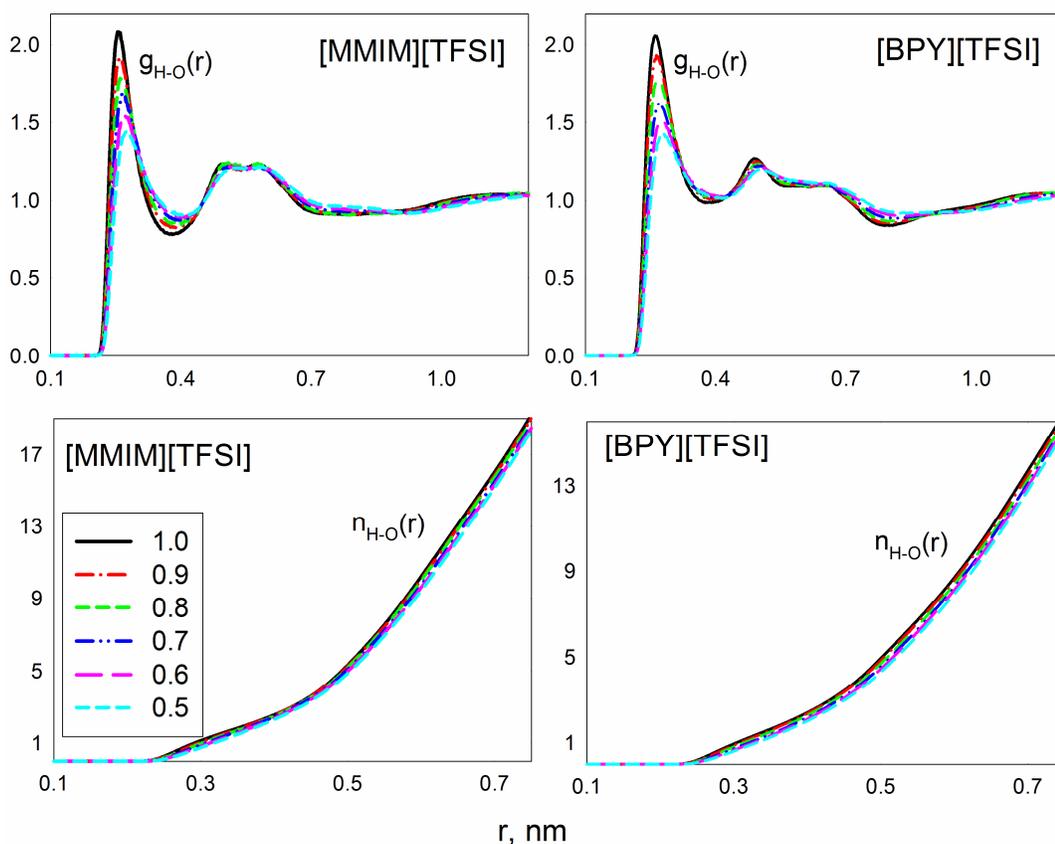

Figure 7. The simulated radial distribution functions and cumulative numbers of (a) [MMIM][TFSI] ($g_{HO}(r)$, $n_{HO}(r)$) and (b) [BPY][TFSI] ($g_{NO}(r)$, $n_{NO}(r)$) with the scaling factors from 1.0 down to 0.5 at 303 K. See legend for shape and color of the lines.

Certainly, a complete neglect of Coulombic interactions is not relevant from the physical point of view. Nevertheless, it provides important insights into the nature of ionic liquids. Table 2 lists the diffusion constants and shear viscosities for [MMIM][TFSI] and [BPY][TFSI] at 303, 353, and 403 K. At the absence of the electrostatic charges, ionic conductivity equals to zero. Remarkably, our results are generally consistent with those previously reported by Borodin and Smith[61], even though they are for another RTIL. The observed behavior once again underlines that Coulombic interactions play a crucial role in determining the nature and behavior of these compounds.

Table 2. The self-diffusion constants of the cation, $D_+$, and anion, $D_-$, and shear viscosity, $\eta$, calculated for [MMIM]$^0$[TFSI]$^0$ and [BPY]$^0$[TFSI]$^0$. Note, the Coulombic interactions are completely turned off for this hypothetic force field model

| Temperature, K | $D_+$, $10^{-11}$ m$^2$/s | $D_-$, $10^{-11}$ m$^2$/s | $\eta$, cP |
|---|---|---|---|
| [MMIM][TFSI] | | | |
| 303 | 96 | 60 | 1.6 |
| 353 | 204 | 102 | 0.6 |
| 403 | 334 | 211 | 0.4 |
| [BPY][TFSI] | | | |
| 303 | 92 | 57 | 1.1 |
| 353 | 171 | 111 | 0.6 |
| 403 | 285 | 210 | 0.5 |

Although fixed-charge FF is not a perfect model from the physical point of view, we illustrate that it is relevant to simulate all physical chemical properties of RTILs realistically. Certainly, different atoms of RTILs polarize in different ways. This phenomenon is not accounted for by means of the proposed methodology. However, the simulation results prove that the impact of varying polarizabilities is nearly negligible as compared to the conventional uncertainties of the simulated data. Note that simulated ions contain a rich set of elements (hydrogen, carbon, nitrogen, oxygen, sulfur, and fluorine), representing five groups and three periods of the periodic table. On a related note, all properties are quite close to ones reported by Borodin[52] who used a more physically relevant approach to carry out a computer study.

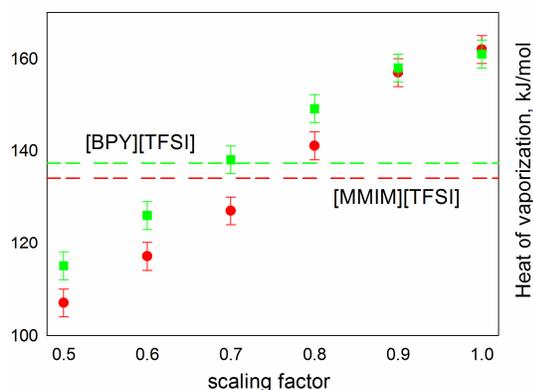

Figure 8. The simulated heats of vaporization of [MMIM][TFSI] (red circles) and [BPY][TFSI] (green squares) with the scaling factors from 1.0 down to 0.5 at 303 K. The dashed horizontal lines of the same color indicate experimental values. The standard deviation of the simulated heat of vaporization is 1-2%.

In addition, the decrease of electrostatic potential around each ion presumably alters bonded force constants which were originally fitted for isolated ions. Therefore, further high level *ab initio* calculations of the ionic clusters are required to clarify this point. In the context of the present work, it should be noted that bonded interaction constants usually play a noticeably less significant role in determining physical chemical properties of the medium than non-bonded interactions do. In order to estimate possible impact of the bonded parameters, we increase and decrease all the original torsional force constants for [MMIM][TFSI] and [BPY][TFSI] by 25%. Remarkably, no observable changes of any of the studied properties are detected. This underlines once again that using certain constraints as implemented in the existing force fields (e.g. eliminating the vibrations of the carbon-hydrogen bonds), is not critical to get reliable results.

**Force Field Generation Proposal**

The best-performing scaling factors, obtained from interpolation of the simulation results, are summarized in Table 1.

Clearly, $f_{sc}$ is unique for each RTIL, but is rather close for all of their thermodynamics, structure and transport properties. Importantly, the variation of $f_{sc}$ with temperature is negligible. Thus, it is possible to refine the original FF by finding only the value of $f_{sc}$. Among other important properties of RTILs, density is probably the most affordable one, which, in addition, can be very fast derived from computer simulation. Whereas a longer trajectory (1-2 ns) is required to properly equilibrate the ionic liquid system at room temperature, the relaxation can be achieved much more quickly at higher (T > 400 K) ones. It is reasonable to start the FF refinement with calculating a specific density of the system for a number of decreasing scaling factors, and obtain the best-performing one using linear interpolation. As the current study demonstrates, the obtained $f_{sc}$ can be successfully applied to simulate other properties. In addition, shear viscosity and heat of vaporization can be applied along with density, if their experimental values are reliable.

A number of the practically important RTILs contain hydrophobic alkyl tails of the varying length. These tails, most likely, do not participate, or participate to a much lesser extent, in the electronic polarization phenomena occurring in the condensed phase of the ionic liquids. They are electrically neutral on the whole, and the charges on carbon and hydrogen atoms are standardized in all general purpose FFs. In the present work, a single $f_{sc}$ is applied to all interaction sites for generality. However, the electrostatics scaling on the alkyl tail is probably excessive. It is even more remarkable that this particular, pretty rough, procedure provides a reasonably good description of both RTILs. This fact underlines that the electrostatic interactions brought by alkyl groups are significantly weaker than the total inter-ionic interactions. They play a minor role in the considered properties which integrally depend on the nature of the ions. Table 3 lists diffusion constants, ionic conductivity, and shear viscosity at 303, 353, and 403 K provided that $f_{sc}$ is not applied to the hydrophobic alkyl tail of [BPY][TFSI]. All the resulting transport properties are within one standard deviation as compared to the uniformly scaled model. At once, note a possibility that the discussed effect can be more pronounced as the alkyl tail becomes longer. Therefore, the Coulombic energy which corresponds to the neutral tails should remain the same as the original FF provides. Although the charges on the alkyl tails should not be scaled from a general physical perspective, this approximation does not bring any systematic errors to the results for [BPY][TFSI].

Table 3. The diffusion constants, $D_+$ and $D_-$, shear viscosity, $\eta$, and ionic conductivity, $\sigma$, of [BPY]$^{+0.7}$[TFSI]$^{-0.7}$, provided that the electrostatic charges on the carbon and hydrogen interaction sites of the alkyl tail are the same as in the original Lopes' force field model

| Temperature, K | $D_+$, $10^{-11}$ m$^2$/s | $D_-$, $10^{-11}$ m$^2$/s | $\eta$, cP | $\sigma$, S/m |
|---|---|---|---|---|
| 303 | 8.2 | 6.5 | 16 | 0.57 |
| 353 | 24 | 19 | 4.3 | 1.7 |
| 403 | 50 | 40 | 2.0 | 3.2 |

The derived methodology is extremely simple and intuitive. It is proven to work well with the Lopes' comprehensive force field for two RTILs, which represent imidazolium and pyridinium families of ionic liquids. The force field of Lopes uses a general transparent procedure of derivation and is available for most of the currently known ionic liquids. Since this set of parameters systematically overestimates densities and underestimate transport properties of the simulated RTILs, $f_{sc} < 1$ is implied. The best-performing $f_{sc}$ for imidazolium and pyridinium RTILs lies between 0.7 and 0.8. The derivation of $f_{sc}$ using the adjustment of density, allows to avoid more computationally demanded procedure, which employs preliminary *ab initio* computation[56].

**Conclusion**

The impact of Coulombic interaction energy on thermodynamics, structure and transport properties of two (imidazolium- and pyridinium-based) RTILs is studied by means of the molecular dynamics simulations employing non-polarizable FF. It is shown that excessive Coulombic energy prevents the simulated ions to exhibit realistic transport properties. The uniformly scaled electrostatic charges with a factor of 0.9, 0.8, 0.7, 0.6, and 0.5 lead to the decrease of Coulombic energy by 19, 36, 51, 64, and 75%, respectively. The scaling of Coulombic energy drastically modifies transport properties of RTILs. Meanwhile, pair distribution functions are almost insensitive to such refinement, and the impact on density and heat of vaporization is far from being crucial. The best-performing scaling factors are characteristic for each simulated RTIL. They do not explicitly depend either on temperature or any of the simulated properties. In spite of the previous failures, it is shown that Lopes' FF can successfully simulate transport properties of RTILs using an extremely simple refinement. The specific density is easily obtainable from experiment. As well, it can be promptly derived from computer simulations. Here, we suggest to exploit this property in order to find the best-performing scaling factor for each RTIL. The refined FF models can be used to successfully simulate all sets of ionic liquids properties.

**Acknowledgment**

It is a pleasure for me to thank Dr. Heather Jaeger (University of Rochester) for a number of valuable hints. The research is supported in part by the NSF grant CHE-1050405.